\documentstyle[12pt,equations,epsf]{article}

\begin{document}
\newcommand{\be}{\begin{equation}}
\newcommand{\ben}{\begin{subequations}}
\newcommand{\een}{\end{subequations}}
\newcommand{\beq}{\begin{eqalignno}}
\newcommand{\eeq}{\end{eqalignno}}
\newcommand{\ee}{\end{equation}}
\newcommand{\tanb}{\mbox{$\tan \! \beta$}}
\renewcommand{\thefootnote}{\fnsymbol{footnote}}

\def\cntwo{\wt\chi_2^0}
\def\wt{\widetilde}
\def\gl{\wt g}
\def\mgl{m_{\gl}}
\def\lsim{\mathrel{\raise.3ex\hbox{$<$\kern-.75em\lower1ex\hbox{$\sim$}}}}
\def\gsim{\mathrel{\raise.3ex\hbox{$>$\kern-.75em\lower1ex\hbox{$\sim$}}}}
\def\fbi{~{\rm fb}^{-1}}
\def\etmiss{/ \hskip-6pt E_T \hskip6pt}
\def\cmone{\tilde{\chi}_1^-}
\def\mcpmone{m_{\cpmone}}
\def\cpone{\tilde{\chi}_1^+}
\def\cpmone{\tilde{\chi}_1^\pm}
\def\cnone{\tilde{\chi}_1^0}
\def\mcnone{m_{\cnone}}
\def\gev{~{\rm GeV}}
\def\mev{~{\rm MeV}}
\def\dmchi{\Delta m_{\tilde{\chi}_1}}
\def\epem{e^+e^-}
\def\anti{\overline}
\def\gam{\gamma}

\pagestyle{empty}
\font\fortssbx=cmssbx10 scaled \magstep2
\hbox to \hsize{
$\vcenter{
\hbox{\fortssbx University of California - Davis}\medskip
%\hbox{\fortssbx University of Wisconsin - Madison}
}$
\vspace*{1.2cm}
$\vcenter{
\hbox{\bf UCD-99-2} 
\hbox{\bf hep-ph/9902309}
\hbox{Feburary, 1999}
}$
}
%}

\begin{center}
{\large\bf
Addendum/Erratum for:\\
`Searching for Invisible and Almost Invisible Particles at
$\bf e^+e^-$ Colliders' [hep-ph/9512230] \\ and \\ 
`A Non-Standard String/SUSY Scenario and its Phenomenological Implications'
[hep-ph/9607421]
\\}
\rm
\vskip2pc
{\bf C.-H. Chen$^a$, M. Drees$^{a,b}$, and J.F. Gunion$^a$}\\
\medskip
\small\it
$^a$Davis Institute for High Energy Physics, 
University of California at Davis,\\
Davis, CA 95616, USA\\
$^b$Inst. de Fisica Teorica, Univ. Estadual Paulista, Sao Paulo, Brazil
\\
\end{center}
\vskip .5cm
\begin{abstract}
We correct our treatment of decays of the lightest chargino
to final states containing the lightest neutralino
in the case where the chargino and neutralino masses differ
by less than 1 GeV. A brief summary of the phenomenological
implications is given.
\end{abstract}

\vspace*{1.0cm}
\renewcommand{\theequation}{A.\arabic{equation}}
\setcounter{equation}{1}
 
Our treatment of exclusive hadronic chargino decays in Refs.~\cite{cdg1,cdg2}
is incorrect. In particular, decays into final states containing an
odd number of pions are not suppressed. Correct expressions for the
corresponding partial widths are given below; these replace and
complete the incorrect Eq.~(A2) in the Appendix of \cite{cdg2}.
\ben \label{en2} \beq
\Gamma(\tilde{\chi}_1^- \rightarrow \tilde{\chi}_1^0 \pi^-)
&= \frac {f_\pi^2 G_F^2} {4 \pi} \frac {|\vec{k}_\pi|}{\widetilde{m}_-^2}
\left\{ \left( O^L_{11} + O^R_{11} \right)^2 \left[ \left(
\widetilde{m}_-^2 - \widetilde{m}_0^2 \right)^2 - m^2_\pi \left(
\widetilde{m}_- - \widetilde{m}_0 \right)^2 \right]
\right. \nonumber \\ & \left.
\hspace*{25mm} + \left( O^L_{11} - O^R_{11} \right)^2 \left[ \left(
\widetilde{m}_-^2 - \widetilde{m}_0^2 \right)^2 - m^2_\pi \left(
\widetilde{m}_- + \widetilde{m}_0 \right)^2 \right] \right\}; \\
\label{en2a}
\Gamma(\tilde{\chi}_1^- \rightarrow \tilde{\chi}_1^0 \pi^- \pi^0)
&= \frac {G_F^2} {192 \pi^3 \widetilde{m}_-^3} 
\int_{4 m_\pi^2}^{(\Delta m_{\tilde{\chi}_1})^2} d q^2 \left| F(q^2) \right|^2
\left( 1 - \frac {4 m_\pi^2}{q^2} \right)^{3/2}
\lambda^{1/2}(\widetilde{m}_-^2,\widetilde{m}_0^2,q^2)
\nonumber \\ & \hspace*{25mm}
\left\{ \left[ \left( O^L_{11} \right)^2 + \left( O^R_{11} \right)^2 \right]
\left[ q^2 \left( \widetilde{m}_-^2 + \widetilde{m}_0^2 - 2 q^2 \right)
+ \left( \widetilde{m}_-^2 - \widetilde{m}_0^2 \right)^2 \right]
\right. \nonumber \\ & \left. \hspace*{25mm}
- 12 O^L_{11} O^R_{11} q^2 \widetilde{m}_- \widetilde{m}_0 \right\}; \\
\label{en2b}
\Gamma(\tilde{\chi}_1^- \rightarrow \tilde{\chi}_1^0 3\pi)
&= \frac {G_F^2} {6912 \pi^5 \widetilde{m}_-^3 f_\pi^2} 
\int_{9 m_\pi^2}^{(\Delta m_{\tilde{\chi}_1})^2} d q^2 
\lambda^{1/2}(\widetilde{m}_-^2,\widetilde{m}_0^2,q^2)
\left| BW_a(q^2) \right|^2 g(q^2)
\nonumber \\ & \hspace*{28mm}
\Bigg\{ \left[ \left( O^L_{11} \right)^2 + \left( O^R_{11} \right)^2 \right]
\left[ \widetilde{m}_-^2 + \widetilde{m}_0^2 - 2 q^2
+ \frac {\left( \widetilde{m}_-^2 - \widetilde{m}_0^2 \right)^2} {q^2} \right]
\nonumber \\ & \hspace*{28mm}
- 12 O^L_{11} O^R_{11} \widetilde{m}_- \widetilde{m}_0 \Bigg\}.
\label{en2c}
\eeq \een 
We have used the same notation as in \cite{cdg2}. $\vec{k}_\pi
= \lambda^{1/2}(\widetilde{m}_-^2, \widetilde{m}_0^2, m^2_\pi)/(2
\widetilde{m}_-)$ in Eq.~(\ref{en2}a) is the pion's 3--momentum in the
chargino rest frame, and $f_\pi \simeq 93$ MeV is the pion decay
constant. The form factor $F(q^2)$ appearing in Eq.~(\ref{en2}b) has
been defined in Eqs.~(A3) and (A4) of \cite{cdg2}.  Explicit expressions
for the Breit--Wigner propagator $BW_a$ of the $a_2$ meson, the exchange of
which is assumed to dominate $3\pi$ production, as well as for the
three-pion phase space factor
$g(q^2)$ can be found in Eqs.~(3.16)-(3.18) of Ref.~\cite{ks}.  
We have used the
propagator without ``dispersive correction''. This underestimates the
partial width for $\tau^- \rightarrow 3 \pi \nu_\tau$ decays by about
35\%; in our numerical results, we have therefore multiplied the
r.h.s. of Eq.~(\ref{en2c}) by 1.35. Nevertheless the branching ratio
for these modes never exceeds $\sim 18\%$; note that Eq.~(\ref{en2c})
includes both $\pi^- \pi^0 \pi^0$ and $\pi^- \pi^- \pi^+$ modes, which
occur with equal frequency.

\setcounter{figure}{0}

\begin{figure}[h]
\leavevmode
\epsfxsize=6.0in
\centerline{\epsffile{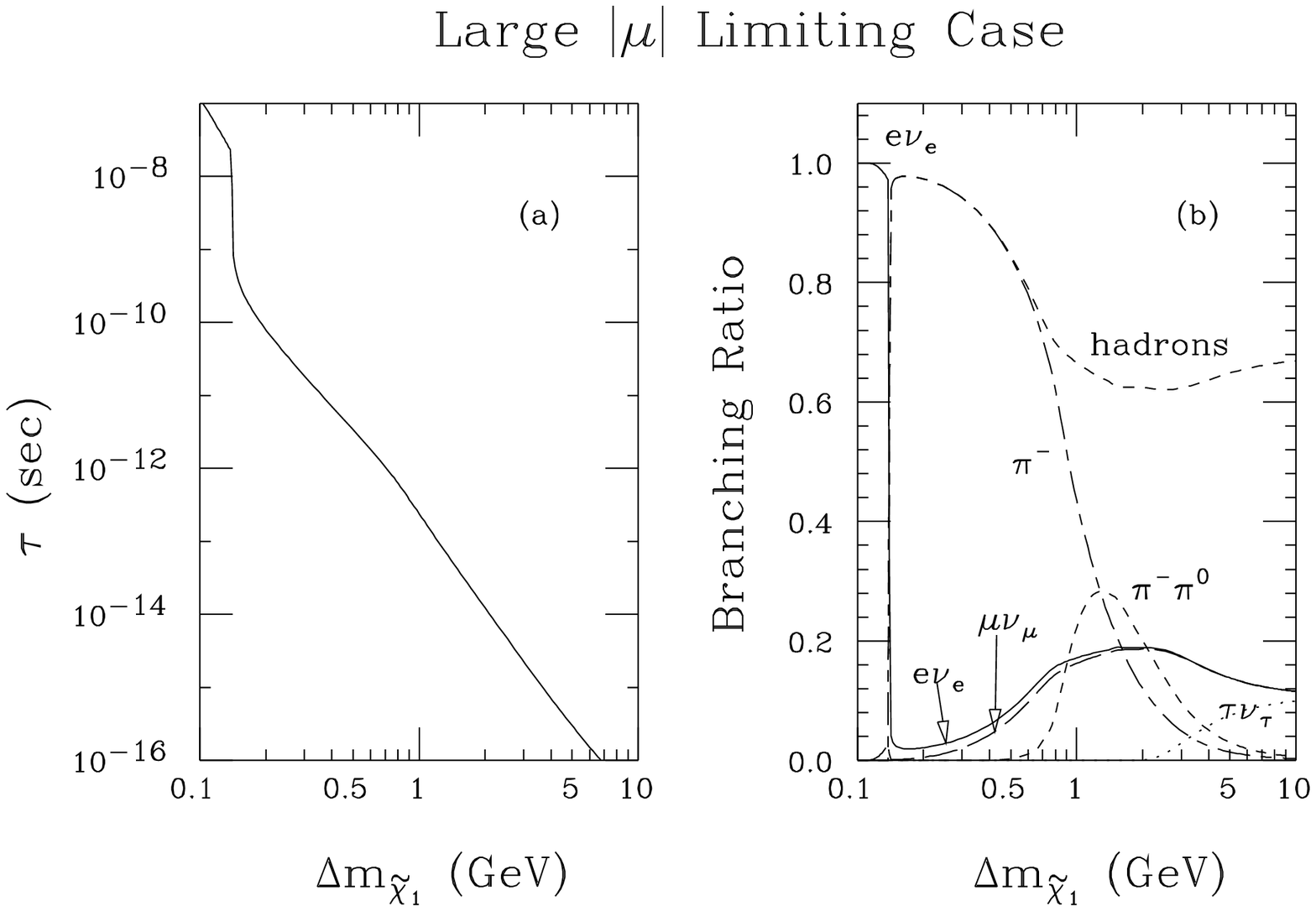}}
\smallskip
\caption
{In (a), we show the lifetime of 
$\tilde{\chi}_1^-$ for the case $M_1 \simeq M_2 \ll
|\mu|$. $\Delta m_{\tilde{\chi}_1}$ is the chargino--neutralino mass
difference. In (b), we give the corresponding
branching ratios of $\tilde{\chi}_1^-$. 
For $\Delta m_{\tilde{\chi}_1} \leq 1.5$ GeV, the branching
ratio for ``hadronic'' decays is computed as the sum of the
branching ratios for 1, 2 and 3 pion final states, while for larger
mass splittings the parton model result has been used.}
\label{lifebrsnew}
\end{figure}

The new results for the lifetime $\tau$ and for the branching ratios
of the $\cmone$ are shown in Fig.~\ref{lifebrsnew}, which is the replacement
for Fig.~2 of Ref.~\cite{cdg1} and Fig.~6 of Ref.~\cite{cdg2}.
The main modifications of our previous results are due to the greatly
increased partial width of the single pion $\cmone\to\cnone\pi^-$ mode
(which now swamps the $\cnone \pi^-\pi^0$ two-pion mode). 
Fig.~\ref{lifebrsnew}(b) shows that the $\cnone\pi^-$ mode
dominates $\tilde{\chi}_1^-$ decays for $m_\pi \leq \Delta m_{\tilde{\chi}_1}
\leq 1$ GeV. As shown in Fig.~\ref{lifebrsnew}(a),
this leads to a rapid drop of the lifetime of the $\cmone$
once this mode opens up. The implications 
for observing chargino pair production at an $\epem$ collider
are the following.

\bigskip
\noindent (a) For $\dmchi\equiv \mcpmone-\mcnone<m_\pi$ our previous results
are unchanged. The produced charginos travel distances of order
a meter or more and appear as heavily-ionizing
tracks in the vertex detector and the main detector, thereby making 
$\epem\to\cpone\cmone$ production background-free
and observable for $\mcpmone$ up to near $\sqrt s/2$.

\bigskip
\noindent (b) For $m_\pi<\dmchi<1\gev$, 
$c\tau$ is smaller than in our original calculation. 
One must consider this range in some detail.
We give a discussion~\footnote{We were greatly aided
in putting this discussion together by
several conversations with H. Frisch from the CDF collaboration.}
appropriate to a silicon vertex detector of the type
planned by CDF for RunII.~\footnote{The vertex detector for
the NLC could be built with similar characteristics. For example,
the Snowmass '96 report {\it Physics and Technology of the
Next Linear Collider}, SLAC Report 485, discusses a 3-5 layer
CCD vertex detector beginning no further out than 2 cm.
Calculated backgrounds at 2 cm are described as `sufficiently
low for efficient vertexing', but do depend upon details
of the interaction region design. Apparently (R. Van Kooten, private
communication), a layer close to $r=1$ cm is being studied.
The SLD vertex detector has an innermost layer at $r=2.5~$cm.  However, the
innermost layers of the vertex detectors at LEP are at
6.3 cm.  Thus the LEP detectors have less ability to see direct evidence
for the $\cpmone$ track for the $c\tau$
range being considered.} This detector will have layers at 
$r\sim 11$, $8.5$, $7$, $4.5$, $3$ and $1.6~\mbox{cm}$
(this latter being the L00 layer) \cite{cdfextra}. 
For $m_\pi<\dmchi<160\mev$, $c\tau>7~\mbox{cm}$, implying that
a $\cmone$ or $\cpone$ produced with low rapidity
will typically pass through 4 or more
layers of the vertex detector before decaying 
(for $\langle\beta\rangle\gsim 0.7$).
This is probably sufficient to recognize
the $\cpmone$ track as being clearly heavily ionizing.
For $160\mev<\dmchi<190\mev$, $7~\mbox{cm}>c\tau>3~\mbox{cm}$
and the $\cpmone$ will typically pass through at least two layers.
Even though these layers would register passage of a heavily-ionizing 
object, this alone might not be enough to clearly identify an unusual event.
However, the $\cpmone$ track will end (which possibly helps
to distinguish it from longer tracks etc. that happen to have
large deposits in the inner few layers) and emit a single
charged pion. The single pion will typically have
transverse momentum of order its momentum,
$p_\pi\sim  \sqrt{\dmchi^2-m_\pi^2}$, in the $\cpmone$ rest frame.
For $160\mev<\dmchi<190\mev$, $p_\pi\sim 77-130\mev$.
The corresponding impact parameter resolution (taking $p_\pi^T\sim p_\pi$),
$b_{\rm res}\sim 300-170~\mu$m (these are the $1\sigma$ values from 
Fig.~2.2 of \cite{cdfextra} when L00 is included),
is much smaller than the actual impact
parameter ($\sim c\tau>3~\mbox{cm}$). Perhaps the combination
of a track that produces large deposits
in a few layers and then ends with the emission of such a pion
will be sufficient to pick out this type of event when combined with
an appropriate trigger. For $\dmchi>230\mev$, $c\tau<1.6~\mbox{cm}$
and the $\cpmone$ will not even pass through the innermost layer
unless it has a very large $\beta$. However, $p_\pi>180\mev$ and
the impact parameter resolution for the single emitted pion
moves into the $<150~\mu$m range.
For example, if $\dmchi=240,300,500,1000\mev$, $c\tau\sim 1.2,
0.37,0.09,0.007~\mbox{cm}$ while $p_\pi\sim 195,265,480,990$ yields
$1\sigma$ impact parameter resolutions of $\sim 120,90,50,25~\mu$m. 
For $\dmchi<1\gev$, we have $c\tau/b_{\rm res}> 3$. We think it
is possible that an event defined by an appropriate trigger
and the presence of one or more high-$b$ charged pions would
be quite distinctive. It seems probable that directly triggering on
$\epem\to \cpone\cmone$ production using just these vertex
detector tracks would be highly problematical. However,
for $m_\pi<\dmchi<1\gev$, 
$\epem\to \gam\cpone\cmone$ would produce an event with an energetic
photon and substantial missing energy in association with either
heavily-ionizing vertex detector tracks or 
charged pions with clearly non-zero impact parameter. We are hopeful
that such events would prove to be essentially background free.
As long as efficiencies for singling out such events
are not very small (a detailed study is required), rates at LEP2 and the NLC
would be adequate for discovery for $\mcpmone$ up to near $\sqrt s/2$.

\bigskip
\noindent (c) For $\dmchi$ above $1\gev$, there
is little change relative to our previous results.
The $\cpmone\to\ell\nu\cnone$ branching ratio is typically $>10\%$ for
$\ell=e$ or $\mu$, with the remainder of the $\cpmone$
decays being into soft multiple pion statess or jets, and $c\tau$ is such that 
the $\cpmone$ decay would be prompt.
As before, if the $\cpmone$ decay is prompt
and if $\dmchi$ is small, the main concern is that
the decay products (hadrons or $\ell\nu$) produced along with the $\cnone$
would be too soft to be distinctively visible in the main part
of the detector. If this is the case, one will have to detect
$\epem\to\gam\cpone\cmone\to \gam+\etmiss$ as an excess
relative to the large $\epem\to\gam Z^*\to\gam \nu\anti\nu$ background.
The resulting limits on the $\mcpmone$ values 
for which the $\gam\cpone\cmone$ signal can be detected are those given
in the the original version of the paper: LEP2 will not improve
the LEP1 $Z$-pole limits on $\mcpmone$ ($\mcpmone<45\gev$)
but the NLC (with $L=50\fbi$)
could probe up to $\mcpmone\sim 200\gev$. A detailed simulation
is required to determine exactly how large $\dmchi$ needs to be in 
order for the soft $\ell$'s or hadrons to be sufficiently energetic to be
detected and direct $\cpone\cmone$ production events identified.
In particular, the boost from the $\cpmone$ rest frame
is determined by both the machine energy and the chargino mass
in question. The strength of the magnetic field
is also important. Studies by the DELPHI group at LEP~\cite{delphideg}
have shown that for $\dmchi>2\gev$ they can pick out the soft leptons
with sufficient efficiency to exclude $\mcpmone<90\gev$ if the
chargino pair cross section is maximal (large sneutrino mass).
Perhaps with a careful design the NLC detectors might be able to
close the gap between this case and case (b) altogether.

\bigskip
\noindent We now give a corresponding summary for a hadron collider.

\bigskip
\noindent (a)
Long-lived heavily-ionizing
tracks from $\cpmone$'s produced in SUSY pair production events
will only be present for $\dmchi<m_\pi$.  For such $\dmchi$
values, events containing a $\cpmone$ will be essentially background
free, and high SUSY mass scales can be probed
using (for example) $\gl\gl$ events in which $\gl\to q^\prime \anti q \cpmone$.
(Note that if the momentum of the long-lived $\cpmone$ is correctly
measured, such events could be reconstructed in such a way that
there is no missing energy associated with the $\cpmone$ decay.)

\bigskip
\noindent (b) For $m_\pi<\dmchi\lsim 1\gev$,
the background to jets + missing energy events in which
gluinos (and squarks) decay to one or more $\cpmone$ might be
largely eliminated assuming that the short heavily-ionizing tracks 
or the pions with non-zero impact parameter coming from
the $\cpmone$'s can be detected in the vertex detector. 
The backgrounds that remain after the usual jet and missing energy cuts
and that are due to multiple scattering in the silicon (causing
low-$b$ pions to apparently have larger $b$),
charged kaon (delayed) decays, and the like, would have to be carefully
studied before a definitive conclusion can be reached.
However, as stated in the original
paper, the standard tri-lepton signal from $\cpmone\cntwo$ production
(that arises when 
$\cpmone\to \ell^{\pm}\nu\cnone$ and $\cntwo\to \ell^+\ell^-\cnone$)
and the like-sign di-lepton signal for $\gl\gl$
production (arising when {\it both} $\gl$'s decay via
$\gl\to q^\prime q\cpone$ followed by $\cpone\to\ell^+\nu\cnone$,
or the charge conjugate) will both be essentially unobservable because of
the softness of the $\ell$'s produced in $\cpmone\to\ell\nu\cnone$
decays.~\footnote{Also, for a bino-like $\cntwo$, the production rate
for $\cpmone\cntwo$ is suppressed.}

\bigskip
\noindent (c) The results of the original
paper obtained for values of $\dmchi$ large enough that the $\cpmone$
decay is prompt, but still too small for the $\cpmone$
decay products to be clearly separable (using
the main tracker and calorimeters of the detector) from soft debris produced
during a typical collision, remain unchanged.
The only mode yielding a viable signal for $\gl\gl$ production (for example)
would be jets + missing energy, and for parameters such
that $\mgl$ is near $\mcpmone$ the discovery reach at the Tevatron
is substantially reduced compared to mSUGRA boundary conditions.

\bigskip
\noindent Please note that
we have replaced the original versions of hep-ph/9512230 and hep-ph/9607421
stored at xxx.lanl.gov with revised versions reflecting the above changes.

\subsection*{Acknowledgements}
We thank James Wells for bringing the problem in our original calculation
to our attention.

\noindent

\end{document}